\documentclass[preprint]{aastex}

\begin{document}

\title{Compton cooling in the afterglows of gamma-ray bursts: \\
Application to GRB 980923 and GRB 971214}

\author{C.-I. Bj\"ornsson}
\affil{Stockholm Observatory, S-133 36, Sweden}
\email{bjornsson@astro.su.se}

\begin{abstract}

The role of Compton cooling in the standard model for the afterglows of 
gamma-ray bursts is considered. When electrons cool by scattering off their own 
synchrotron photons, three cooling regimes are identified in which the observed 
synchrotron radiation exhibits qualitatively different characteristics. 
Depending on the values of the source parameters, an afterglow may 
evolve through one, two or all three of these regimes. Since 
synchrotron radiation can be regarded as Compton scattering of the 
virtual photons due to the magnetic field, in one of these regimes 
the instantaneous synchrotron spectrum has properties identical to those when Compton 
cooling is negligible. During this phase in the evolution, the synchrotron radiation falls 
mainly in the near infrared/optical/UV spectral range. In order to 
break this degeneracy, good temporal coverage is needed.
Alternatively, the importance of Compton scattering can be determined 
for those afterglows which are observed outside this 
degeneracy phase. It is suggested that the afterglows of GRB 980923 and GRB 971214 are 
two such cases. In the prompt afterglow of GRB 980923, the 
Klein-Nishina limit suppresses Compton scattering and the cooling is 
due to synchrotron radiation. However, the derived values of the source 
parameters are such that Compton cooling is expected to have been 
important in its subsequent evolution. The observed properties of GRB 
971214 indicate cooling to be dominated by Compton scattering rather 
than synchrotron radiation. If Compton cooling is generally important 
in the afterglows of gamma-ray bursts, the likelihood of the shock 
becoming radiative is increased. It is suggested that this effect
contributes to the low frequency of detected afterglows.

\end{abstract}


\section{Introduction}

In view of its relative simplicity, the standard model for the 
afterglows of gamma-ray bursts \citep{M/R97, SPN98} has proven to be surprisingly 
successful \citep[e.g.,][]{WRM97, Gal98}. Synchrotron radiation from behind a spherically 
symmetric shock, caused by the interaction between an external 
medium and a relativistically expanding fireball, accounts for an 
impressive amount of observed characteristics. Deviations from the 
expected behavior observed for some afterglows during their later 
stages have been attributed to an aspherical outflow, for example, a 
jet geometry. Jet-edge effects and the lateral expansion of the outflow 
then increase the rate of 
decline of the light curves as well as the typical synchrotron 
frequency \citep{Roa97, Har99}. However, afterglows exist for which the later 
stages of the evolution show no signs of deviation from a spherical 
outflow \citep[e.g.,][]{Fru99}.

\citet{Kul99} noticed 
several afterglows to have soft X-ray fluxes in excess of 
that obtained by extrapolating from the infrared/optical spectral range. 
Such an additional spectral component is not expected and 
indicates that the standard model is incomplete. There are also some 
afterglows whose observed properties in the infrared/optical spectral 
range do not comply with standard predictions \citep[e.g.,][]{W/G99}. 
Although the properties of most observed afterglows are well 
accommodated within the standard model, the reason is not clear for the small 
likelihood of a given gamma-ray burst to produce an observed 
afterglow; for example, an adiabatic outflow requires
rather large variations in the source parameters between different 
gamma-ray bursts 
in order to suppress the emitted radiation below the observed values for 
the majority of the afterglows. This issue has been discussed 
by \citet{Ake00} for the prompt afterglows. We suggest these 
shortcomings to be due, at least in part, to the neglect of Compton cooling. In order 
not to increase the number of free parameters in the standard model, 
it is assumed that Compton scattering occurs on the self-produced 
synchrotron photons. 

Normally, arguments used for not considering 
Compton scattering together with synchrotron radiation include: 
The observations are well explained by synchrotron cooling only and 
energy equipartition between relativistic electrons and magnetic 
fields makes Compton scattering marginally important, at most, for the 
observed spectral evolution of the afterglows \citep{SNP96, S/P99, MSB00}. For many 
afterglows, observations are not extensive enough to allow a 
determination of independent values for the source parameters and, 
hence, the assumption of equipartition is one of convenience, since 
the plasma properties behind the shock are not understood well enough 
to strongly motivate such a choice. However, sometimes the preferred value for 
the partition of energy between electrons and magnetic field is such 
that a consistent description requires Compton scattering to be 
included \citep[e.g.,][]{C/L99, Vre99}. This is in line with the arguments of
\citet{Gal99} that a major 
reason for the varied appearance of afterglows is a wide range of 
values for the magnetic field. The effects on the synchrotron 
spectrum from first order Compton scattering have been discussed 
by \citep{P/K00, S/E00, W/L00}. 

Synchrotron radiation can be seen as Compton scattering of virtual 
photons due to the magnetic field. Hence, in a situation when all 
electrons scatter off the same real photons, the instantaneous 
synchrotron spectrum is indistinguishable from one for which the 
cooling is due to synchrotron radiation \citep[see also][]{S/E00}. The 
main aim of the present paper is three-fold: (i) Determine the phase 
of the afterglow evolution during which this spectral degeneracy 
occurs. (ii) Characterize the spectral properties and their evolution 
outside this degenerate phase for situations when Compton cooling 
affects the afterglow. (iii) Apply these results to observed 
afterglows in order to estimate the role played by Compton scattering.

There are two suggested physical settings for the gamma-ray burst 
phenomenon, namely, the merger of a neutron star with another compact object 
and the explosion of a hypernova \citep{Pac98}. For the afterglow, the main difference 
between these two scenarios is the assumed properties of the 
external medium into which the shock moves. In the former case, the external medium is assumed to 
have properties similar to the interstellar medium, while in the 
latter case the shock encounters a Wolf-Rayet--type stellar wind. The importance of Compton 
scattering is determined mainly by the value of the energy 
density in electrons relative that in magnetic fields. Since this 
ratio is likely due to local plasma processes behind the shock, 
rather than global properties, the inclusion of Compton scattering 
is qualitatively similar for the two cases.
Hence, for convenience, the discussion is limited to the case with an 
interstellar--type external medium. 

In \S\,2 the relevant timescales 
are defined for afterglows in which the cooling is dominated by 
Compton scattering. It is shown that the evolution can contain up to 
three phases each with its own cooling characteristics. The spectral 
properties, including light curves, of these phases are discussed in \S\,3. 
Two specific afterglows (GRB 980923 and GRB 971214) are discussed in 
\S\,4. They have been chosen since their observed properties suggest 
them to be outside the degenerate phase. The values derived for the 
source parameters imply Compton cooling to have significantly affected 
the evolution of both these afterglows. The implications for 
afterglows in general are briefly discussed in \S\,5.

\section{Cooling behind the shock: Compton scattering versus 
synchrotron radiation}

In the standard model for the afterglow of gamma-ray 
bursts, electrons and magnetic fields are injected through a relativistic 
shock. There is evidence for a non-spherical shock; however, as long 
as the Lorentz factor of the shock is larger than the inverse of the 
angular extent of the asphericity, the observed evolution is well 
described by a spherically symmetric shock. The calculations below 
assume the validity of such a description. Furthermore, the discussion 
is restricted to the forward shock.

\subsection{The importance of Compton cooling}

We will adopt the usual convention of measuring the thermodynamic 
quantities of the shocked gas, including the energy density of photons, 
in its own restframe, while other quantities are 
measured in the restframe of the external medium.
Let the electron energy distribution, injected behind the spherically symmetric 
shock, be given by $N(\gamma)\propto \gamma^{-p}$ for 
$\gamma\geq\gamma_{m}$ and zero otherwise, where $\gamma$ is the 
Lorentz factor of the electron. The typical 
frequency emitted by an electron with Lorentz factor 
$\gamma_{m}$ is $\nu_{m}$ and $\nu_{c}$ is 
the frequency emitted by those electrons ($\gamma_{c}$), which have a 
cooling time equal to the dynamical time scale of the shock. In order 
to simplify the discussion, $2<p<3$ is assumed so that the 
peak of the energy density in synchrotron photons occurs at a frequency 
$\nu\approx\max\left(\nu_{m},\nu_{c}\right).$  

When cooling is complete, the energy density of photons behind a 
relativistic shock is roughly equal to the energy density injected 
through the shock. 
If cooling is not complete (i.e., 
$\gamma_{c}>\gamma_{m}$), the photon energy density is smaller by a factor 
$\approx(\gamma_{c}/\gamma_{m})^{2-p}$. Therefore, the energy density 
of synchrotron photons, $U_{synch}$, at $\nu_{c}$ is given by
\begin{equation}
	\frac{U_{synch}(\nu_{c})}{B^{2}/8\pi}
	\approx\frac{\epsilon_{e}}{\epsilon_{B}}\left(\frac
	{\gamma_{c}}{\gamma_{m}}\right)^{2-p},
	\label{2.1}
\end{equation}
where $\epsilon_{e}$ and $\epsilon_{B}$ are the 
fractions of the injected energy which go into relativistic electrons 
and magnetic fields ($B$), respectively. The modification to 
equation (\ref{2.1}), needed when Compton scattering off the self-produced 
synchrotron photons dominates the cooling, can be obtained as follows. 
Since synchrotron radiation can be regarded as Compton scattering of 
virtual photons due to the magnetic field (`zeroth' order Compton 
scattering), 
$U_{synch}(\nu)\approx\gamma^{2}
\tau(\gamma)B^{2}/8\pi$, where 
$\tau(\gamma)$ is the Thomson optical depth of 
electrons, with Lorentz factor $\sim\gamma$, radiating at $\nu$. 
Likewise, when Compton scattering dominates the cooling, the main 
cooling of electrons with Lorentz factor $\gamma$ occurs on photons 
with frequency $\nu_{\gamma}$ with a corresponding energy density
$U_{Comp}(\nu_{\gamma})$. Therefore, the analogue of 
equation (\ref{2.1}), when Compton scattering dominates synchrotron radiation as 
the main cooling mechanism, takes the form
\begin{equation}
	\frac{U_{synch}(\nu_{c})}{B^{2}
	/8\pi}\frac{U_{Comp}(\nu_{\gamma_{c}})}{B^{2}
	/8\pi}\approx\frac{\epsilon_{e}}{\epsilon_{B}}\left(\frac
	{\gamma_{c}}{\gamma_{m}}\right)^{2-p}.
	\label{2.2} 
\end{equation}\\

The relation given in equation (\ref{2.2}) is a formal one and the expressions 
for $U_{Comp}$ and $\nu_{\gamma}$ have to be determined for 
each situation. There are two simple, limiting cases; scattering in the 
Klein-Nishina limit which implies $\nu_{\gamma}\propto\gamma^{-1}$ 
and scattering in the Thomson limit when $\nu_{\gamma}$ is 
independent of $\gamma$ and all electrons cool on the 
same photons. The latter
case can arise, for example, when $\gamma_{m}$ is large enough
for the emission in first order Compton scattering to be well separated from 
the synchrotron component while, at the same time, the Klein-Nishina limit prevents 
appreciable radiation in second order Compton scattering. In such a 
situation $U_{Comp}=U_{synch}$ and 
$\nu_{\gamma}=\max(\nu_{m},\nu_{c})$ and equation (\ref{2.2}) reduces 
to 
\begin{equation}
	\frac{U_{synch}(\nu_{c})}{B^{2}
	/8\pi}\approx\left[\frac{\epsilon_{e}}{\epsilon_{B}}
	\left(\frac
	{\gamma_{c}}{\gamma_{m}}\right)^{2-p}
	\right]^\frac{1}{2}.
	\label{2.3}
\end{equation}
Cooling due to scattering in the Thomson limit is expected to be similar to 
synchrotron cooling, since, as mentioned above, synchrotron 
radiation can be regarded as `zeroth' order Compton scattering on 
virtual photons, which are the same for all electrons independent of 
their energy.
On the other hand, cooling in the 
Klein-Nishina limit should exhibit characteristics quite different 
from those of either synchrotron cooling or scattering in the Thomson limit.
We now discuss and contrast the salient features of the observed radiation in 
these three limits, i.e., cooling due to synchrotron radiation, 
Compton scattering, respectively, in the Thomson and Klein-Nishina limits.

\subsection{Characteristics of afterglows when cooling is dominated by Compton 
scattering}

The above arguments show that Compton cooling is important whenever 
\begin{equation}
	\frac{\epsilon_{e}}{\epsilon_{B}}\geq \zeta \, \mbox{max}\left[1,
	\left(\frac{\nu_{c}}{\nu_{m}}\right)^{
	\frac{p-2}{2}}\right].
	\label{2.4}
\end{equation}
The value of $\zeta$ is model dependent but $\zeta \sim v/c$ is 
expected, where $v$ is the velocity of the shock relative the shocked 
gas. Hence, as mentioned above, for a relativistic shock $\zeta$ is of order unity.
When source parameters are such that the condition in equation 
(\ref{2.4}) is 
fulfilled, Compton scattering should affect the evolutionary 
properties of 
the afterglow. It is convenient to divide the observed behavior 
into three phases according to the different cooling regimes. Synchrotron cooling dominates early on, 
while Compton scattering in the Thomson limit can take over during an 
intermediate stage. For $\gamma_{m}$ small enough, multiple Compton 
orders become possible. Hence, at late times multiple scattering 
may be possible in which cooling is likely to occur in the Klein-Nishina 
limit. Furthermore, cooling in the Klein-Nishina limit is expected to 
be important also during the transition between the different phases. 
These separate phases will be referred to as the synchrotron, 
Thomson, and multiple Compton scattering phase, respectively.
As is shown below, in the standard scenario for the 
afterglow, its emission may go through one, two, or all three of these 
phases.   

The scattering cross-section for an electron with Lorentz factor $\gamma$ drops 
rapidly for photon frequencies larger than $\nu_{KN}(\gamma)$ (the 
Klein-Nishina limit), given by
\begin{equation}
 	\gamma\frac{\nu_{KN}(\gamma)}{\Gamma}\approx\frac{
 	mc^2}{h},
 	\label{2.5}
\end{equation}
where $\Gamma$ is the 
bulk Lorentz factor of the shocked gas, $m$ is the electron mass, and 
$h$ is the Planck constant. 
Following \citet{SPN98}, we use 
$\gamma_{m}=6.1\times10^{2}\epsilon_{e}\Gamma$, which corresponds to 
$p=2.5$ and define 
$\eta\equiv\gamma/\gamma_{m}$ or, equivalently, $\nu\equiv\eta^2\nu_{m}$, 
which leads to
\begin{equation}
	\nu_{KN}(\eta)\approx\frac{2.0\times 10^{17}}
	{\epsilon_{e}\eta}
	\label{2.6}
\end{equation}\\
The expressions for $\nu_{m}$ and $\nu_{sc}$, the 
cooling frequency for negligible Compton losses, can be written as
\begin{equation}
	\nu_{m}=1.2\times10^{15}E^{1/2}_{52}t^{-3/2}_{d}
	\epsilon^{2}_{e}\epsilon^{1/2}_{B}
	\label{2.7}
\end{equation}
and
\begin{equation}
    \nu_{sc}=2.9\times 10^{12}E^{-1/2}_{52}n^{-1}
    t^{-1/2}_{d}\epsilon^{-3/2}_{B},
	\label{2.8}
\end{equation}
where $E_{52}$ is the total energy of the shocked gas in units of
$10^{52}$ ergs, $t_{d}$ is the observed time, in days, since the onset 
of the expansion, and $n$ is the number density of the external 
medium. Furthermore, $\Gamma =6.7(E_{52}/n)^{1/8}t_{d}^{-3/8}$ 
\citep[henceforth WG99]{W/G99} has been used in equations 
(\ref{2.7}) and (\ref{2.8}). The 
cooling frequency, including the effects of Compton 
scattering, can then be written
\begin{equation}
    \nu_{c}=\frac{\nu_{sc}}{\left[1+U_{Comp}(\nu_{\gamma_{c}})
    /(B^2/8\pi)\right]^2}.
	\label{2.9}
\end{equation}

The time dependence of these frequencies can be used to determine the 
characteristic times when the salient features of the Compton scattering  
change. The properties of the observed synchrotron radiation depend 
sensitively on the relative values of $\nu_{m}$ and $\nu_{c}$. 
Consider the case when scattering occurs in the Thomson limit and 
let $t_{eq}$ denote the time when $\nu_{m}=\nu_{c}$. With the use of 
$U_{Comp}(\nu_{\gamma_{c}}) \approx U_{synch}(\nu_{c})
\approx (\epsilon_{e}/\epsilon_{B})^{1/2}(B^2/8\pi)$ in 
equation (\ref{2.9}), one finds for $t_{eq}$, in days, 
\begin{equation}
 	t_{d,eq}=4.1\times10^2 E_{52}n\epsilon_{e}^3\epsilon_{B}.
 	\label{2.10}
\end{equation}
With the use of equation (\ref{2.3}), the time dependence of 
$\eta_{c}\equiv (\nu_{c}/\nu_{m})^{1/2}$ can then be written
\begin{equation}
	\eta_{c}=\left(\frac{t}{t_{eq}}\right)^{1/2}\mbox{max}\left\{1,\left(
	\frac{t}{t_{eq}}\right)^{(p-2)/[2(4-p)]}\right\}.
	\label{2.11}
\end{equation}
Note that this expression for $\eta_{c}$ is generally valid for 
Compton scattering in the Thomson limit, i.e., it is not restricted to 
the case when $t=t_{eq}$  falls in the Thomson phase. 

\begin{sloppypar}
Assume that $\eta_{c}<1$ at the beginning of the Thomson phase. An 
electron with Lorentz factor $\gamma$ scatters in the Thomson 
limit when $\nu_{KN}(\eta)>\nu_{m}$. From equations (\ref{2.6}) and 
(\ref{2.7}) 
this leads to
\begin{equation}
  t(\eta)>t_{1C}\eta^{2/3}
  \label{2.12}
\end{equation}
where the value of $t_{1C}$, in days, is
\begin{equation}
  	t_{d,1C}=3.3\times 10^{-2}E_{52}^{1/3}\epsilon_{e}^{2}
  	\epsilon_{B}^{1/3}.
  	\label{2.13}
\end{equation}
Hence, $t_{1C}$ marks the transition from the synchrotron to the 
Thomson phase and the condition $\eta_{c}<1$ at the transition is equivalent to 
$t_{eq}>t_{1C}$. If instead $\eta_{c}>1$ at the beginning of the 
Thomson phase, the condition for scattering in the Thomson limit becomes 
$\nu_{KN}(\eta)>\nu_{c}$ or
\begin{equation}
  t(\eta)>t_{1C}\left(\frac{\eta}{\eta_{c}}\right)^{2/3}\eta_{c}^2.
  \label{2.14}
\end{equation}
However, since $\eta_{c}^2$ increases more rapidly with time than 
linearly in the Thomson phase for $\eta_{c}>1$ ($2<p<3$), equation 
(\ref{2.14}) leads to a
contradiction, since $\eta\geq\eta_{c}$ for cooling to be important. 
As a result, $t_{eq}<t_{1C}$ implies that no transition 
to the Thomson phase takes place and the source remains in the 
synchrotron phase during the entire evolution. The reason is that the value 
of $\gamma_{c}$ increases with time so 
fast that the Klein-Nishina limit suppresses efficient cooling due to 
Compton scattering.
\end{sloppypar}

The condition for the source to remain in the Thomson phase for 
$t>t_{eq}$ (i.e., $\eta_{c}>1$) is
\begin{equation}
 	\gamma_{c}^{2}\nu_{c}>\nu_{KN}(\eta)
 	\label{2.15}
\end{equation}
This leads to
\begin{equation}
 	t(\eta)<t_{2C}\left(\frac{\eta}{\eta_{c}}\right)^{4/9}\eta_{c}^{20/9},
 	\label{2.16}
\end{equation}
where the value of $t_{2C}$, in days, is
\begin{equation}
 	t_{d,2C}=1.6\times10^{2}E_{52}^{1/3}n^{-1/9}
 	\epsilon_{e}^{20/9}\epsilon_{B}^{2/9}.
 	\label{2.17}
\end{equation}
Again, if the condition in equation (\ref{2.16}) is valid at $t=t_{eq}$, it 
remains so for all times, since $\eta_{c}^{20/9}$ increases more rapidly with time than 
linearly. Hence, for $t_{eq}<t_{2C}$, no transition occurs to 
the multiple Compton scattering phase. In order for such a transition 
to be possible, $t_{eq}>t_{2C}$ is required, in which case the 
transition takes place at $t=t_{2C}$ (cf. eq. [\ref{2.16}] with 
$\eta_{c}=1$). 

When there is no transition to the multiple Compton scattering phase, 
the source remains in the Thomson phase as long as
$\nu_{KN}(\eta_{c})>\nu_{c}$, which implies
\begin{equation}
	t<t_{eq}\left(\frac{t_{eq}}{t_{1C}}\right)^{(4-p)/(p-2)}.
	\label{2.18}
\end{equation}
Furthermore, in order for Compton cooling to be important the 
inequality in equation (\ref{2.4}) must be valid, which results in 
($\zeta=1$)
\begin{equation}
	t<t_{eq}\left(\frac{\epsilon_{e}}{\epsilon_{B}}\right)^{(4-p)/(p-2)}.
	\label{2.19}
\end{equation}
Hence, for $t_{1C}<t_{eq}<t_{2C}$, the duration of the Thomson phase is determined by the 
relative values of $t_{eq}/t_{1C}$ and $\epsilon_{e}/\epsilon_{B}$. 
These characteristic times are useful for describing the properties of 
the emitted radiation only as long as $t\lesssim t_{d}(\Gamma = 1)$, where 
$t_{d}(\Gamma =1) = 1.6\times 10^2 (E_{52}/n)^{1/3}$, in days, is the 
approximate time when the outflow becomes subrelativistic.

The values of $t_{1C}$ and $t_{2C}$ correspond to the times when those electrons, 
which provide the main contribution to the photon energy density, are 
able to scatter their own radiation into 
the next higher Compton order (cf. eqs. [\ref{2.12}] and [\ref{2.16}]). However, as 
mentioned above, 
cooling in the next higher Compton order may become important before 
these transition times. The scattering then occurs in the 
Klein-Nishina limit on frequencies below that where the photon energy 
density peaks. The extent of these transition stages depends 
on the relative strength of the Compton orders (i.e., 
$\epsilon_{e}/\epsilon_{B}$).

With the use of 
equations (\ref{2.10}), (\ref{2.13}), and (\ref{2.17}) the relation $t_{1C}<t_{eq}<t_{2C}$ 
can be written 
\begin{equation}
	 8.0\times10^{-5}<E_{52}^{2/3}n\epsilon_{e}\epsilon_{B}^{2/3}<3.4
	 \times10^{-1}n^{-1/9}\epsilon_{e}^{2/9}\epsilon_{B}^{-1/9}. 
	\label{2.20}
\end{equation}
An alternative way of writing this relation is
\begin{equation}
	7.2\times 
	10^{-7}E_{52}^{-1}n^{-3/2}\epsilon_{e}^{-5/2}<\epsilon_{B}/\epsilon_{e}<
	2.5\times 10^{-1}E_{52}^{-6/7}n^{-10/7}\epsilon_{e}^{-2}.
	\label{2.20b}
\end{equation}
The three cooling regimes in Figure 1 are delimited by the three 
characteristic times $t_{1C}$, $t_{2C}$, and $t_{eq}$. Energy 
requirements suggest a value of $\epsilon_{e}$ not much smaller than 
unity. Hence, since $t_{eq}$ is more sensitive than $t_{1C}$ and 
$t_{2C}$ to variations in $E_{52}$ and $n$ (cf. eqs. [\ref{2.10}], 
[\ref{2.13}], and [\ref{2.17}]), the change which occurs for values 
of $E_{52}$ and $n$ other than the ones used in Figure 1 ($E_{52}=10$ 
and $n=1$) is due mainly to the variation in $t_{eq}$ ($\propto E_{52}n$). 
It is seen from equation (\ref{2.20b}) and Figure 1 that a significant 
multiple Compton scattering phase can occur during the later stages of 
the afterglow for large values of $E_{52}$ and/or $n$. 
Likewise, small values of $E_{52}$ and $n$ limit the extent of the 
Thomson phase or can exclude it altogether for small values of 
$\epsilon_{B}/\epsilon_{e}$. Although $\epsilon_{B}/\epsilon_{e}\ll1$ 
for such cases, synchrotron radiation dominates the cooling due to 
the Klein-Nishina limit in the scattering cross-section.

In summary (see Figure 1), the value of $t_{eq}$ relative to $t_{1C}$ 
and $t_{2C}$ determines the salient features of the Compton scattered 
radiation.
If Compton cooling significantly affects the observed emission, 
a rough characterization of the source evolution can be 
done as follows.\\
$t_{eq}<t_{1C}$: {\it Synchrotron phase}. Initially, cooling is dominated by synchrotron 
radiation. Later on first order Compton scattering in the 
Klein-Nishina limit may become important but no transition to the 
Thomson phase takes place.\\
$t_{1C}<t_{eq}<t_{2C}$: {\it Synchrotron and Thomson phases}. 
Like the previous case but here a transition 
occurs to the Thomson phase at 
$t\approx t_{1C}$. 
A transition back to the synchrotron phase 
takes place for values of $t_{eq}$ close to $t_{1C}$ (cf. eq. 
[\ref{2.18}]), while values of $t_{eq}$ close to $t_{2C}$ can cause 
second order Compton scattering in the Klein-Nishina limit 
to become important later on.\\
$t_{2C}<t_{eq}$: {\it Synchrotron, Thomson, and multiple Compton scattering phases}. Like the 
previous case except that at $t\approx t_{2C}$ multiple Compton 
scatterings become possible. Cooling is then likely to occur in the Klein-Nishina 
limit. 

\section{The emitted synchrotron spectrum when Compton cooling is 
important}

Depending on the values of the source parameters (cf. eq. 
[\ref{2.20}] or [\ref{2.20b}]), 
the Compton 
scattering can evolve through one, two, or all three of the phases 
discussed above. The observed characteristics of the resulting 
synchrotron radiation are quite different in the various phases 
and we now consider them in turn.

\subsection{The synchrotron phase}

In the beginning of this phase, cooling is dominated by synchrotron 
radiation. At some later time, depending on the value of 
$\epsilon_{B}/\epsilon_{e}$, first order Compton scattering in the 
Klein-Nishina limit may take over as the main cooling agent. For this 
latter stage of the synchrotron phase it is convenient to 
introduce a parameter $\eta_{crit}$; 
it corresponds to the normalized Lorentz factor of those electrons, which can Compton 
scatter on their own synchrotron photons in the Klein-Nishina limit 
(cf. eq. [\ref{2.6}]),
\begin{equation}
	\eta_{crit}=\left(\frac{2.0\times 10^{17}}{\epsilon_{e}\nu_{m}}
	\right)^{1/3}.
	\label{3.1}
\end{equation} 
Consider first the case $\eta_{crit}<1$. This condition holds during 
the whole synchrotron phase for $t_{eq}>t_{1C}$. Assuming no cooling
at $\nu_{m}$ leads to $U_{Comp}(\nu_{\eta}) \propto 
\eta^{-4/3}$ for $\eta\geq 1$. This shows the Compton cooling time 
scale to decrease with decreasing electron energy. Hence, if 
significant Compton 
cooling occurs in the synchrotron phase for $\eta_{crit}<1$, 
electrons radiating at $\nu_{m}$ are affected. For such situations
the spectral shape below $\nu_{m}$ is important. The minimum value of 
$\eta$, for which 
the main Compton cooling occurs on photons below $\nu_{m}$ is obtained 
from $\nu_{KN}(\eta_{min})\equiv\nu_{m}$ or
\begin{equation}
 	\eta_{min}=\frac{2.0\times 10^{17}}{\epsilon_{e}\nu_{m}}=\eta_{
 	crit}^3
 	\label{3.2}
\end{equation}
The corresponding synchrotron frequency is
\begin{equation}
  	\nu_{min}=\eta_{min}^2\nu_{m}=\eta_{crit}
  	^6\nu_{m}
  	\label{3.3}
\end{equation}
The electrons which cool in the Klein-Nishina limit on frequencies 
around $\nu_{min}$ have a normalized Lorentz factor, $\eta_{max}$, 
determined from $\nu_{KN}(\eta_{max})\equiv\nu_{min}$ or
\begin{equation}
	\eta_{max}=\frac{2.0\times 10^{17}}{\epsilon_{e}\nu_{min}}
	=\eta_{crit}^{-3}.
	\label{3.4}
\end{equation}
The corresponding synchrotron frequency is
\begin{equation}
	\nu_{max}=\eta_{max}^2\nu_{m}=\eta_{crit}
  	^{-6}\nu_{m}.
	\label{3.5}
\end{equation}
Hence, electrons with 
$\eta_{crit}^3<\eta<\eta_{crit}^{-3}$ cool on 
photons in the frequency range $\nu_{m}\eta_{crit}^{6}<\nu<\nu_{m}$. If Compton 
cooling is important for these electrons, their energy distribution 
can be obtained in a self-consistent way.

Let $N(\eta)\propto\eta^{-p_{c}}$ for $\eta_{crit}^3<\eta<1$. 
In the corresponding frequency range, the spectral flux, 
$F_{\nu}\propto\nu^{-\alpha}$, has $\alpha=(p_{c}-1)/2$. 
Furthermore, steady injection of electrons with $\eta \geq 1$
implies $N(\eta)\propto
-(d\eta/dt)^{-1}$. Since it is assumed that Compton scattering 
dominates the cooling $d\eta/dt\propto -\eta^2 U_{Comp}(\nu_{\eta})
\propto -\eta^{(p_{c}+1)/2}$, 
where $\nu_{\eta}\propto\eta^{-1}$. Equating these two independent 
expressions for $N(\eta)$ yields $\eta^{p_{c}}=\eta^{(p_{c}+1)/2}$,
which implies $p_{c}=1$ and, hence, $\alpha=0$ in the frequency range 
$\eta_{crit}^6<\nu/\nu_{m}<1$. Furthermore, electrons radiating 
in the frequency range $1<\nu/\nu_{m}<\eta_{crit}^{-6}$ suffer 
cooling at a rate $d\eta/dt\propto{-\eta}$. This results in an exponentially 
decreasing energy with time, i.e., cooling time is independent of 
electron energy. As a consequence, in the latter frequency range, 
cooling does not affect the spectral index, i.e., $\alpha=\alpha_{o}
\equiv(p-1)/2$. Hence, the spectral break at 
$\eta_{crit}^{-6}\nu_{m}$ could be mistaken for a cooling break. 
However, since $\eta_{crit}^{-6} \propto \nu_{m}^2 \propto t^{-3}$, 
such a spectral break would decrease with time much faster than 
any cooling frequency. The actual cooling break occurs below 
$\eta_{crit}^{6}\nu_{m}$.

\begin{sloppypar}
Although of less interest for the present paper, the rest of the 
spectrum can be determined in a similar manner (Fig. 2). Electrons radiating in 
the frequency range below 
$\eta_{crit}^6 \nu_{m}$ cool on photons with frequencies larger 
than $\nu_{m}$. Unless $\eta_{crit}$ is close to unity, cooling of these
electrons occurs on photons in the $\alpha_{o}$ part of the 
spectrum. This results in a steepening of the spectrum below 
$\eta_{crit}^{6} \nu_{m}$ from $\alpha=0$ to $\alpha=\alpha_{o}/2$. 
The high energy  
electrons, which cool on photons in the $\alpha_{o}/2$ part of the 
spectrum, radiate at frequencies larger than $\eta_{crit}^{-6}\nu_{m}$. With 
the same reasoning as above, this causes a steepening of the 
spectrum from $\alpha=\alpha_{o}$ 
to $\alpha=(5/4)\alpha_{o}$. At even higher frequencies, the radiating 
electrons cool on the low frequency, $F_{\nu}\propto\nu^{1/3}$, 
part of the synchrotron spectrum. This cooling, if it is 
important, results in a flattening of the spectrum and the spectral index is 
given by $\alpha =\alpha_{o}-1/6$. At these high frequencies, 
synchrotron cooling can become important even if 
$\epsilon_{e}\gg\epsilon_{B}$ because the effective value of 
$U_{Comp}$ decreases rapidly with increasing $\gamma$.
\end{sloppypar}

\begin{sloppypar}
In the synchrotron phase, $\eta_{crit}>1$ occurs for $t_{eq}<t_{1C}$ 
and implies $\nu_{c}>\nu_{m}$. 
Electrons with $\eta_{crit}^{-3}<\eta<\eta_{crit}^{3}$ have 
$\nu_{m}<\nu_{KN}(\eta)<\eta_{crit}^{6}\nu_{m}$. If Compton cooling is 
important and $\nu_{c}<\eta_{crit}^{6}\nu_{m}$, 
$\nu_{KN}(\eta_{c})>\nu_{m}$ is implied, which results in 
$U_{Comp}(\nu_{\eta_{c}})/(B^2/8\pi)\propto\nu_{m}^{(p-1)/8}\nu_{c}^{(p-5)/8}$ 
(cf. eq. [\ref{2.2}]). 
Inserting this in equation (\ref{2.9}) yields $\nu_{c}\propto t^{(3p-7)/[2(p-1)]}$. 
The observed value of $p$ is such that $\nu_{c}$ should 
increase with time for most afterglows. An even faster increase of $\nu_{c}$ with time 
occurs for $\nu_{c}>\eta_{crit}^{6}\nu_{m}$, since then 
the cooling takes place on the $U_{synch}\propto \nu^{4/3}$ part of 
the spectrum.
\end{sloppypar}

The above discussion shows that the spectrum in the synchrotron 
phase is expected to 
consist of a rather large number of spectral ranges with different 
spectral indices. The finite spectral width of the emission from the 
individual electrons causes a gradual transition from one spectral 
range to another. Unless a given spectral range is rather wide, the 
spectral indices will vary continuously with frequency and the 
expressions  
derived above should be regarded as a rough guide to the typical values 
expected in the different parts of the spectrum. Furthermore,
this smoothing of the 
spectral features is increased by the finite escape time of the photons 
from behind the relativistic shock.

\subsection{The Thomson phase} 

The Thomson phase occurs for $t_{eq}>t_{1C}$ and begins at $t\approx 
t_{1C}$, which corresponds to 
\begin{eqnarray}
     \nu_{m} & = & 2.0\times 10^{17}\epsilon_{e}^{-1}
     \label{3.6}\\
     \nu_{c} & = & 1.6\times10^{13}E_{52}^{-2/3}n^{-1}\epsilon_{e}^{-2}
     \epsilon_{B}^{-2/3}\left(=\nu_{m}\frac{t_{1C}}{t_{eq}}\right).
     \label{3.7}
\end{eqnarray}
In this phase the Klein-Nishina limit plays a minor role, except 
preventing significant amount of second order Compton scattering. 
Since all the electrons cool on the same photons, i.e., 
$\nu_{\gamma}=\max(\nu_{m},\nu_{c})$ does not depend on $\gamma$, 
the behavior of the synchrotron flux is expected to be similar to the case
when only 
synchrotron radiation contributes to the cooling. The main 
differences which arise from using $U_{Comp}$ instead of 
$B^{2}/8\pi$ in the cooling rates, concern the magnitude and time 
dependence of $\nu_{c}$ \citep[see also][]{P/K00}.

As shown in \S\,2.2 (see also Fig. 1), $t<t_{eq}$ in the Thomson phase for $t_{eq}>t_{2C}$.
For such conditions, the light curves during the Thomson 
phase are the same as those obtained for negligible cooling 
due to Compton scattering (the synchrotron case). The only difference is 
the value of $\nu_{c}$, which is smaller by a factor 
$\epsilon_{B}/\epsilon_{e}$ as compared to the value in the synchrotron 
case. 
The situation changes for $t_{eq}<t_{2C}$, since then $t=t_{eq}$ occurs 
during the Thomson phase and the steeper time dependence of $U_{Comp}$ 
for $t>t_{eq}$ 
causes the rate at which $\nu_{c}$ declines to slow down. Since 
$\nu_{m}\propto t^{-3/2}$, the value 
of $\nu_{c}$ even increases for $p>8/3$ (cf. eq. [\ref{2.11}]). The change in 
the 
time dependence of $\nu_{c}$ takes place when $\nu_{c}=\nu_{m}\equiv 
\nu_{eq}$, where
\begin{equation}
	\nu_{eq}=1.4\times10^{11}E_{52}^{-1}n^{-3/2}
	\epsilon_{e}^{-5/2}\epsilon_{B}^{-1}.
	\label{3.8}
\end{equation}
This causes the light curves 
to differ from the synchrotron case for $t>t_{eq}$. Let the light curves be
characterized by $\beta$ such that the flux $F_{\nu}\propto t^{-\beta}$.
In the synchrotron case, there is a 
break in the light curves as $\nu_{c}(>\nu_{m})$ passes through a given 
frequency band ($\nu<\nu_{eq}$) and $\beta$ increases by $1/4$ from $3(p-1)/4$ to 
$(3p-2)/4$ \citep{SPN98}. 
Since $F_{\nu}\propto \nu_{c}^{1/2}$ for $\nu>\nu_{c}$, 
equation (\ref{2.11}) shows this break to be less abrupt in the Thomson 
phase; for $p<8/3$,
the value of $\beta$ increases only by $1/4-(p-2)/[2(4-p)]=(8-3p)/[4(4-p)]$.
In addition to this chromatic break, there is also an achromatic break 
occurring at $t=t_{eq}$ in the 
light curves for $\nu>\nu_{eq}$, which is absent in the synchrotron 
case.
Again using the fact that $F_{\nu}\propto \nu_{c}^{1/2}$ for 
$\nu>\nu_{c}$ together with equation (\ref{2.11}) result in a decrease 
of $\beta$ by (p-2)/[2(4-p)] from its earlier value (3p-2)/4, i.e., this break
corresponds to a flattening of the light curves (Fig. 3a). The value of 
$\nu_{c}$ increases with time for $p>8/3$ and $t>t_{eq}$. In such 
situations, there is no break in the light curves for $\nu < \nu_{eq}$, 
instead there are two breaks for $\nu > \nu_{eq}$; first comes the 
achromatic break and then the chromatic one (Fig. 3b). As was 
mentioned in \S\,1, the instantaneous synchrotron spectral properties in the 
Thomson phase should be identical to those in the synchrotron case. 
It is seen here that this is also true for the light curves when 
$t<t_{eq}$.

The solution for the source parameters in terms of the observables in 
the synchrotron case was given by WG99. The corresponding solution in 
the Thomson phase can be obtained in a similar manner 
\citep[cf.][]{S/E00}. In the Thomson phase one can write for $\nu_{c}>\nu_{m}$
\begin{mathletters}
\begin{eqnarray}
 	F_{\nu_{m},norm} &=& E_{52}n^{1/2}\epsilon_{e}^{1/2} 
 	\label{3.9a}\\
 	\nu_{m,norm} &=& E_{52}^{1/2}\epsilon_{e}^2\epsilon_{B}^{1/2}
 	\label{3.9b}\\
 	\nu_{c,norm} &=& 
 	E_{52}^{-1/2}n^{-1}\epsilon_{e}^{-1}\epsilon_{B}^{-1/2}\left(
 	\frac{\nu_{c}}{\nu_{m}}\right)^{(p-2)/2}
 	\label{3.9c}\\
 	\nu_{a,norm} &=& 
 	E_{52}^{1/5}n^{3/5}\epsilon_{e}^{-1}\epsilon_{B}^{1/5},
 	 \label{3.9d}
\end{eqnarray}
\end{mathletters}
where $F_{\nu_{m},norm}, \nu_{m,norm}, \nu_{c,norm},$ and 
$\nu_{a,norm}$ are the normalized expressions, respectively, for 
$F_{\nu_{m}}$, $\nu_{m}, \nu_{c},$ and the synchrotron self-absorption 
frequency $\nu_{a}$. The normalization is the same as used 
by WG99. From equations (\ref{3.9a}) -- (\ref{3.9d}), the source 
parameters can be expressed in terms of the observables as
\begin{mathletters}
\begin{eqnarray}
 	E_{52} &=& F_{\nu_{m},norm}^2\nu_{m,norm}^{-3/2}\nu_{c,norm}^{-1/2}
 	\nu_{a,norm}^{-5/2}\left(\frac{\nu_{c}}{\nu_{m}}\right)^{(p-2)/4} 
 	\label{3.10a}\\
 	n &=& \nu_{m,norm}^{-7/6}\nu_{c,norm}^{-3/2}
 	\nu_{a,norm}^{-5/6}\left(\frac{\nu_{c}}{\nu_{m}}\right)^{3(p-2)/4} 
 	\label{3.10b}\\
 	\epsilon_{e} &=& \nu_{m,norm}^{-1/6}\nu_{c,norm}^{-1/2}
 	\nu_{a,norm}^{-5/6}\left(\frac{\nu_{c}}{\nu_{m}}\right)^{(p-2)/4} 
 	\label{3.10c}\\
 	\epsilon_{B} &=& 
 	F_{\nu_{m},norm}^{-2}\nu_{m,norm}^{25/6}\nu_{c,norm}^{5/2}
 	\nu_{a,norm}^{35/6}\left(\frac{\nu_{c}}{\nu_{m}}\right)^{-5(p-2)/4}. 
 	\label{3.10d}
\end{eqnarray}
\end{mathletters}
In order for the solution in equation (\ref{3.10a}) -- (\ref{3.10d}) to be consistent, 
$\epsilon_{e}/\epsilon_{B}>\zeta(\nu_{c}/\nu_{m})^{(p-2)/2}$.
From equations (\ref{3.10c}) and (\ref{3.10d}) this 
requirement is equivalent to 
\begin{equation}
	\left(\frac{\nu_{c}}{\nu_{m}}\right)^{p-2}>\zeta
	F_{\nu_{m},norm}^{-2}\nu_{m,norm}^{13/3}\nu_{c,norm}^{3}
 	\nu_{a,norm}^{20/3}.
	\label{3.11}
\end{equation}

A consistent solution for the synchrotron case requires 
$\epsilon_{e}/\epsilon_{B}<\zeta (\nu_{c}/\nu_{m})^{(p-2)/2}$. 
With the use of the equations in WG99, one finds 
\begin{equation}
	\left(\frac{\nu_{c}}{\nu_{m}}\right)^{p-2}>\zeta^{-2}
	F_{\nu_{m},norm}^{-2}\nu_{m,norm}^{13/3}\nu_{c,norm}^{3}
 	\nu_{a,norm}^{20/3}.
	\label{3.12}
\end{equation}
Except for the different $\zeta$-dependence, the consistency 
requirement is the same for the Thomson phase and synchrotron case. 
Since $\zeta\sim 1$ for a relativistic shock, this shows that 
for the afterglows of gamma-ray bursts, 
the two solutions cannot be distinguished from each other when only 
the synchrotron component of the spectrum is observed. It is 
straightforward to show that this conclusion remains valid also for 
the stage when $\nu_{c}<\nu_{m}$. The degeneracy between these two 
solutions will be broken if the derived source parameters do not 
comply with the requirements discussed in \S\,2.2 for being in the 
Thomson phase.

\subsection{The multiple Compton scattering phase}

The multiple Compton scattering phase occurs for $t_{eq}>t_{2C}$ and 
begins at $t\approx t_{2C}$, which corresponds to
\begin{eqnarray}
     \nu_{m} & = & 5.7\times10^{11}n^{1/6}\epsilon_{e}^{-4/3}
     \epsilon_{B}^{1/6}
     \label{3.13}\\
     \nu_{c} & = & 
     2.3\times10^{11}E_{52}^{-2/3}n^{-17/18}\epsilon_{e}^{-19/9}
     \epsilon_{B}^{-11/18}\left(=\nu_{m}\frac{t_{2C}}{t_{eq}}\right).
     \label{3.14}
\end{eqnarray}
For parameters typical of the afterglows of gamma-ray bursts, at these 
late times the 
different Compton orders are expected to overlap; hence, the scattered 
flux forms a continuum in which the different orders are not easily 
distinguished. The electrons then cools in the Klein-Nishina limit, 
i.e., $\nu_{\gamma}=\nu_{KN}(\gamma)$. The requirement 
that the total energy of the shock should not be excessive suggests 
that the value of $\epsilon_{e}$ is not much smaller than unity.  
Equations (\ref{3.13}) and (\ref{3.14}) then show that the synchrotron radiation in the 
multiple Compton scattering phase is likely to fall mainly in the far 
infrared/radio 
spectral range. The properties of the observed flux under such conditions are 
discussed by \citet{B/A00}.

\section{Discussion}

The best studied afterglows have been observed mainly in the near 
infrared/optical spectral range. For values of source parameters 
thought relevant for the standard model, this corresponds to the 
frequency range where the synchrotron radiation is expected to be 
emitted during the Thomson phase (cf. eqs. [\ref{3.6}] and [\ref{3.13}]). 
In this phase, it is hard to distinguish observationally between two very 
different physical situations; namely, one in which Compton 
scattering dominates the cooling and another where Compton scattering 
plays a minor role. In fact, a rather large parameter space exists (cf. 
eq. [\ref{2.20}] or [\ref{2.20b}]) in which it is not 
even in principle possible to separate the two cases when only the 
synchrotron component is observed \citep[cf.][]{S/E00}.

In this section the afterglows of GRB 980923 and GRB 971214 are 
discussed. They have been selected primarily because 
their observed characteristics suggest that they do not entirely 
radiate in the phase where the values of the source parameters are degenerate. Hence, these 
afterglows should allow a determination of the importance of Compton 
scattering in their evolution. The afterglow of GRB 980923 is likely to 
be in the synchrotron phase. The rather odd 
behavior of the infrared/optical afterglow in GRB 971214 (eg., WG99) 
is ascribed to the source being in the transition stage between 
the Thomson and multiple Compton scattering phases.

\subsection{GRB 980923}

The BATSE observations of the prompt afterglow from GRB 980923 exhibited a 
break in the spectrum which evolved with time towards lower 
frequencies \citep{Gib99}. Interpreted as synchrotron 
radiation from the forward shock, \citet{Gib99} showed this break to 
be consistent with the synchrotron cooling frequency, $\nu_{sc}$,
and incompatible with $\nu_{m}$. Their conclusion is based on two independent 
pieces of evidence, namely the time evolution of the break frequency 
as well as the flux. However, as shown in \S \,3.2, synchrotron radiation and 
Compton scattering in the Thomson phase have similar cooling
characteristics. Since $\nu_{c}>\nu_{m}$, the cooling frequency 
evolves slower with time in the Thomson phase than in the synchrotron case. 
The observed value $p=2.4$ implies $\nu_{c}\propto t^{1/4}$ (cf. 
eq. [\ref{2.11}]) instead 
of $\nu_{sc}\propto t^{1/2}$. Although the observed light curve favors 
synchrotron cooling, the uncertainties in both the flux 
and spectral slope are such that Compton scattering in the Thomson 
phase cannot be 
excluded. We therefore consider both cases together.  

It is convenient to introduce the parameter $\epsilon_{e,crit}\equiv 
\zeta\,\epsilon_{B}(\nu_{c}/\nu_{m})^{(p-2)/2}$; this 
is the value of $\epsilon_{e}$ for which cooling due to synchrotron 
radiation equals that in 
Compton scattering (cf. eq. [\ref{2.4}]). The observed 
fluence ($\mathcal{F}$) can then be related to the source parameters as 
\begin{mathletters}
\begin{eqnarray}
 	\mathcal{F} & = & \frac{\epsilon_{B}E(1+z)}{4\pi 
 	d^2}\frac{\epsilon_{e}}{\epsilon_{e,crit}}, 
 	\label{4.1a}\\
 	\mathcal{F} & = & \frac{\epsilon_{B}E(1+z)}{4\pi 
 	d^2}\left(\frac{\epsilon_{e}}{\epsilon_{e,crit}}\right)^{1/2},
 	\label{4.1b}
\end{eqnarray}
\end{mathletters}
where equations (\ref{4.1a}) and (\ref{4.1b}) correspond, respectively, to cooling 
due to synchrotron radiation and Compton scattering in the Thomson 
phase. The lower fluence in the Thomson phase is due to the fact that 
only a fraction $(\epsilon_{e,crit}/\epsilon_{e})^{1/2}$ of the 
emitted energy emerges as synchrotron radiation (cf. eq. [\ref{2.3}]). 
The luminosity distance to the source is denoted by $d$.
Likewise, expressions for $\epsilon_{B}$ are obtained from 
equations (\ref{2.8}) and (\ref{2.9}). With the use of the observed value $\nu_{c}=3.0\times 
10^{19}$ Hz at $t= 40$ s, one finds
\begin{mathletters}
\begin{eqnarray}
 	\epsilon_{B} & = & 2.7\times 
 	10^{-4}n^{-2/3}E_{52}^{-1/3}(1+z)^{-1/3},
 	\label{4.2a} \\
 	\epsilon_{B} & = & 2.7\times 
 	10^{-4}n^{-2/3}E_{52}^{-1/3}(1+z)^{-1/3}
 	\left(\frac{\epsilon_{e,crit}}{\epsilon_{e}}\right)^{2/3},
 	\label{4.2b}
\end{eqnarray}
\end{mathletters}
for the two cases. The fluence 
in the afterglow for $t>40$ s is $\approx 3.0\times10^{-5}\mbox{ergs 
cm}^{-2}$, which results in
\begin{mathletters}
\begin{eqnarray}
 	E_{52} & = & 1.6\times 10^{6}n (1+z)^{-1}d_{28}^3\left(\frac{ 
 	\epsilon_{e,crit}}{\epsilon_{e}}\right)^{3/2} 
 	\label{4.3a} \\
 	E_{52} & = & 1.6\times 10^{6}n (1+z)^{-1}d_{28}^3\left(\frac{ 
 	\epsilon_{e}}{\epsilon_{e,crit}}\right)^{1/4},
 	\label{4.3b}
\end{eqnarray}
\end{mathletters}
where $d_{28}$ is the luminosity distance in units of $10^{28}$ cm. 
It is seen from equations (\ref{4.3a}) and (\ref{4.3b}) that observations require
$E_{52}>1.6\times 10^{6}n (1+z)^{-1}d_{28}^3$
for both the synchrotron ($\epsilon_{e}<\epsilon_{e,crit}$) and Thomson 
($\epsilon_{e}>\epsilon_{e,crit}$) case. For standard values of 
$E_{52}$ and $n$, this places the source in the local universe
(i.e., $d_{28}\ll 1$). Since the distance to GRB 980923 is not known 
such a solution cannot be ruled out {\it a priori}. The identical 
constraints on the source parameters obtained for the synchrotron 
case and Compton scattering in the Thomson phase is an 
example of the general result derived in \S \,3.2.  

\begin{sloppypar}A cosmological distance 
can be made compatible with standard values for $E_{52}$ and $n$
if the afterglow of GRB 980923 is in the synchrotron phase and 
$\epsilon_{e}\gg\epsilon_{e,crit}$. Assuming the afterglow 
to be in the transition stage between the synchrotron and 
Thomson phases implies that the cooling is due to first order Compton 
scattering in the Klein-Nishina limit. However, as discussed in \S 
\,3.1, possible spectral breaks under such conditions either increase 
with time ($p=2.4$) or decrease too rapidly with time to be 
compatible with the evolution of the observed spectral break. Hence, a 
consistent description requires the cooling to be due to synchrotron 
radiation.

The requirement of negligible Compton cooling at $\nu_{sc}$ constrains the allowed 
values of the source parameters. For a given observed flux at 
$\nu_{sc}$, the value of $U_{Comp}(\nu_{\eta_{sc}})$ increases with 
$\eta_{crit}$; hence, the weakest constraints are obtained for 
$\eta_{crit}<1$. Since no low frequency break in the spectrum is apparent, $\nu_{m}< 
6.0\times 10^{18}$ Hz. The expression for $\eta_{crit}$ (eq. [\ref{3.1}]) 
together with $\eta_{crit}<1$ then 
requires $\epsilon_{e}>1/30$. If Compton cooling is important at 
$\nu_{m}$, the electrons would cool down quickly to at least 
$\eta_{min}=\eta_{crit}^{3}$, which results in a flat ($\alpha = 0$) 
spectrum in the frequency range $\eta_{crit}^{6}\leq \nu/\nu_{m} \leq 1$ 
(cf. \S \,3.1). This increase in flux below $\nu_{m}$ makes the cooling 
time independent of frequency in the range $\eta_{crit}^{6}\leq 
\nu/\nu_{m} \leq \eta_{crit}^{-6}$. Hence, assuming 
$\nu_{sc}<\eta_{crit}^{-6}\nu_{m}$, no Compton cooling at $\nu_{sc}$ 
is equivalent to no Compton cooling at $\nu_{m}$.
\end{sloppypar}

Negligible Compton cooling at $\nu_{m}$ requires
\begin{equation}
	\nu_{m}<\frac{\nu_{sc}}{\left[U_{Comp}(\nu_{\gamma_{m}})
	/(B^2/8\pi)\right]^2}.
	\label{4.4}
\end{equation}
Since the energy density below $\nu_{m}$ is proportional to 
$\nu^{4/3}$, $U_{Comp}(\nu_{\gamma_{m}})=\eta_{crit}^4 
U_{synch}(\nu_{m})$. 
Furthermore, $U_{synch}(\nu_{m})/(B^{2}/8\pi)=(\epsilon_{e}/\epsilon_{B})
(\nu_{m}/\nu_{sc})^{1/2}$ and equation (\ref{4.4}) yields
\begin{equation}
	\frac{\epsilon_{B}}{\epsilon_{e}}> \eta_{crit}^{4}
	\frac{\nu_{m}}{\nu_{sc}}.
	\label{4.5}
\end{equation}
With the use of equation (\ref{3.1}) and 
$\nu_{sc}=3.0\times10^{19}$
this can be written 
\begin{equation}
	\epsilon_{B}> 6.7\times10^{-3}\eta_{crit}.
	\label{4.6}
\end{equation}

When $\nu_{sc}>\eta_{crit}^{-6}\nu_{m}$, the restrictions set by the 
lack of 
Compton scattering is more model dependent. However, 
a definite lower limit to $\epsilon_{B}$ can be obtained by assuming negligible 
Compton cooling only at $\nu_{sc}$ but (unphysically) no enhancement of the 
flux below $\nu_{m}$ due to cooling of electrons with 
$\eta<\eta_{sc}$. The condition 
$U_{Comp}(\nu_{\eta_{sc}})<B^{2}/8\pi$ then leads to
\begin{equation}
	\frac{\epsilon_{B}}{\epsilon_{e}}> \eta_{crit}^{4}
	\left(\frac{\nu_{m}}{\nu_{sc}}\right)^{7/6}.
	\label{4.7}
\end{equation}
This lowers the limit 
of $\epsilon_{B}$ in equation (\ref{4.5}) only marginally. With the use of 
the expression for $\eta_{crit}$ (eq. [\ref{3.1}]) and the observed upper 
limit of $\nu_{m}$, one finds from equation (\ref{4.7}) 
\begin{equation}
	\epsilon_{B}>1.6\times 10^{-3}\epsilon_{e}^{-1/3}
	\label{4.8}
\end{equation}
or a factor $1.3$ larger if equation (\ref{4.5}) is used.

Although the main aim of the present paper is to discuss the 
qualitative features introduced by significant Compton cooling in 
an otherwise standard synchrotron model, it should be mentioned that 
the minimum values of $\epsilon_{B}$ in equations (\ref{4.5}) and 
(\ref{4.7}) are 
likely to be slightly too restrictive for two reasons. The above estimate 
assumes $\zeta=1$. A lower value may be appropriate, since
even for a relativistic shock $v=c/3$. 
Furthermore, the value used for the Klein-Nishina frequency is too
large; the decline in the cross-section sets in at somewhat lower frequencies
than assumed in equation (\ref{2.5}) \citep{B/G70}. For 
example, using $\zeta=1/3$ and a value of the
Klein-Nishina frequency half as large as that in equation (\ref{2.5}) 
reduce the lower limit of 
$\epsilon_{B}$ by a factor 7.6 as compared to those 
given above. 
 
With the assumption of a cosmological distance to GRB 980923, the 
observations of its afterglow together with equations (\ref{4.2a}) 
and (\ref{4.3a}) indicate: (1) A value of $E_{52}$ 
substantially larger than unity and/or a value of $n$ substantially 
smaller than unity, although within the range thought applicable for 
the standard scenario. (2) The value of $\epsilon_{B}$ needs to be 
close to its lower limit allowed by the observed lack of Compton 
cooling ($\sim
10^{-3}$). (3) The value of $\epsilon_{e}$ cannot be much 
smaller than unity. This shows that in 
order for the afterglow to be detected by BATSE, the values of 
$\epsilon_{e}$ and $\epsilon_{B}$ are quite constrained. For values 
of $\epsilon_{B}$ considerably smaller than given in equation 
(\ref{4.8}) Compton 
cooling dominates and for values larger than this, 
$\nu_{sc}$ falls below the BATSE frequency range; in either 
case, the result is a diminishing observed fluence. Likewise, a 
value of $\epsilon_{e}$ considerably smaller than unity also lowers 
the fluence below the detectability of BATSE. Hence, in general, for canonical 
values of $E$ and $n$, the detection of a prompt afterglow by BATSE 
implies values of $\epsilon_{B}$ and $\epsilon_{e}$ not too different 
from the ones derived above for GRB 980923.

\begin{sloppypar}\citet{Sar97} pointed out that for large values of $\epsilon_{e}$ the 
shock may become radiative if $\nu_{c}\leq\nu_{m}$. He showed that the 
energy in the shock decreases, roughly, as $E(t)=E_{o}(t/t_{o})^{-\epsilon_{e}}$.
In the case of GRB 980923, 
$\nu_{c}\geq\nu_{m}$ for most of the synchrotron phase. With the 
parameters derived above $t_{eq}>t_{1C}$ (cf. eq. [\ref{2.20}] or 
[\ref{2.20b}]) and, hence, the afterglow 
should enter the Thomson phase, which implies 
$t_{o}\sim t_{1C}$. The cooling of the shock in the Thomson phase 
continues 
until $t\sim t_{eq}$. The fraction of the initial shock energy 
which is radiated away is then $1-(t_{1C}/t_{eq})^
{\epsilon_{e}}$. The rather large value deduced above for 
$\epsilon_{e}$ suggests that radiative cooling may have been important for the 
dynamics of the afterglow of GRB 980923 in its later phases.
\end{sloppypar}

\subsection{GRB 971214}

The afterglow of GRB 971214 showed a drop in the K-band flux by a 
factor 3--4 between $t(=t_{1}) = 0.20$ days and $t(=t_{2}) = 0.58$ days 
\citep{Gor98}. 
At $t_{2}$ the spectrum exhibited 
a break at $ 3\times10^{14}$ Hz \citep{Kul98, Ram98} together with 
evidence for significant reddening \citep{Hal98}. Within the 
context of a 
standard synchrotron model, this optical break can be 
due either to $\nu_{m}$ or $\nu_{sc}$. However, WG99
emphasized that either choice gives rise to serious inconsistencies. 
Interpreted as $\nu_{m}$, the break predicts an increasing flux in 
the K-band between $t_{1}$ and $t_{2}$ rather than the observed 
decrease. WG99 used the synchrotron model and its expected 
correlation between spectral index and rate of decline of the light curve to correct for 
the reddening. Assuming that the break corresponds to $\nu_{sc}$, and 
$\nu_{m}<\nu_{sc}$, in order to account for the decline of K-band 
flux, the dereddening leads to a 
flat or weakly declining intrinsic flux distribution below the break. 
This is in contrast to the predicted rise corresponding to a spectral index 
$(p-1)/2$. WG99 also discussed several other unattractive 
features, which result from a straightforward application of the 
synchrotron model.

Interpreting the optical break as $\nu_{sc}$ leaves the X-ray flux unexplained, while it is
accounted for when the break is due to $\nu_{m}$. However, the latter 
case strains the derived source parameters; for example, in order to be a 
consistent solution, $\epsilon_{e}/\epsilon_{B}\lesssim 
(\nu_{sc}/\nu_{m})^{(p-2)/2}$ (cf. eq. [\ref{2.4}]). The observed values 
$\nu_{sc}/\nu_{m}\gtrsim 5\times 10^3$ and $p=2.7$ together with 
$\epsilon_{e}\lesssim 
1$, imply $n\lesssim 2\times 10^{-3}$. This upper limit to the density is much smaller than 
values expected in the standard model. Hence, even disregarding the initial 
K-band flux measurements, the observed spectrum at $t_{2}$ makes the 
synchrotron model incomplete.

In view of the limitations of the synchrotron model to account for 
the observed properties of the afterglow in GRB 971214, we suggest 
the discrepancy between observations and model predictions to be due 
to the neglect of Compton scattering. 
Below we discuss a scenario in 
which the observations correspond to the stage when the dominant 
contribution to the cooling changes from first to second order 
Compton scattering. The infrared/optical emission is 
synchrotron radiation, while the X-ray emission is due 
to Compton scattering of the lower frequency synchrotron radiation.
The optical spectral break is assumed to be due to 
$\nu_{m}$, which, in turn, is larger than $\nu_{c}$, i.e., cooling is 
important for all electrons injected at the shockfront. At $t_{1}$ 
second order Compton scattering starts to dominate the cooling at 
$\nu_{K} (\geq\nu_{c})$, the frequency corresponding to the K-band. For larger 
frequencies cooling is still in the Thomson limit. 
Hence, the spectral index between $\nu_{K}$ and $\nu_{m}$ is $1/2$. 
At $t_{2}$ second order Compton scattering dominates the cooling in 
the frequency range below $\nu_{m}$. It is assumed 
that $\nu_{KN}(\gamma_{K})$ lies on the $F_{\nu}\propto\nu^{1/3}$ part of the first 
order Compton component. Hence, 
$U_{Comp}(\nu_{\gamma})\propto\gamma^{-4/3}$, which leads to a
spectral index of $-1/6$ between
$\nu_{K}$ and $\nu_{m}$ (cf. \S\,3.1).

\begin{sloppypar}
With the use of $\nu_{m}\propto t^{-3/2}$ and $F_{\nu_{m}}\propto 
t^{1/2}$ ($t\lesssim t_{2}$), the fluxes at $\nu_{K}$ 
at $t_{1}$ and $t_{2}$, respectively, are related by
\begin{equation}
	\frac{F_{\nu_{K}}(t_{2})}{F_{\nu_{K}}(t_{1})} = 
	\left(\frac{\nu_{K}}{\nu_{m}(t_{2})}\right)^{2/3}\left(\frac{t_{1}}
	{t_{2}}\right)^{1/4}
	\label{4.9}
\end{equation}
Since $\nu_{m}(t_{2})/\nu_{K}\sim 3$ and 
$t_{2}/t_{1}\sim 3$, equation (\ref{4.9}) yields 
$F_{\nu_{K}}(t_{1})/F_{\nu_{K}}(t_{2})\sim 3$, which is consistent 
with the observed value. Hence, the proposed scenario can account 
for the decline in the K-band flux as well as the spectral index below 
the break at $t_{2}$. We now discuss the values of the 
source parameters which result from the above interpretation of the 
observations. The aim is to show that such a scenario is plausible 
rather than to perform a detailed comparison. In order to achieve the 
latter, numerical calculations are necessary; in particular, since 
second order Compton scattering is assumed to play an important role.
\end{sloppypar}

\begin{sloppypar}
Including the redshift dependence in equation (\ref{2.7}), the observed values 
of $\nu_{m}(t_{2}) = 3.0\times 10^{14}$ and $z = 3.42$ lead to
\begin{equation}
	E_{52}\epsilon_{B}\epsilon_{e}^4 = 2.8\times10^{-3}.
	\label{4.10}
\end{equation}
With the use of equation (\ref{4.10}) one finds in the Thomson phase 
\begin{equation}
	\frac{\nu_{c}}{\nu_{m}} = 4.0\times10^{-2}\frac{\epsilon_{e}}{n}
	\frac{t}{t_{1}}.
	\label{4.11}
\end{equation}
It is shown below that a consistent solution suggests a value 
of $\epsilon_{e}/n$ not substantially below unity. The time dependence 
of $\nu_{m}$ implies 
$\nu_{m}(t_{1}) \approx 15 \nu_{K}$. Hence, $\nu_{c}$ in equation 
(\ref{4.11}) is close 
to $\nu_{K}$ at $t_{1}$ and second order Compton scattering affects the 
cooling only for $t\gtrsim t_{1}$. 
Furthermore, the K-band flux should reach a maximum for 
$\nu_{c}\approx \nu_{K}$, i.e., at $t\approx t_{1}$. The observations of 
\citet{Gor98} are consistent with this expectation; although the errors 
are rather large, the flux 
measured at $t=0.15$ is roughly equal to that at $t_{1}$ ($=0.20$), 
i.e., the decline in the K-band flux is likely to have started at 
$t \approx t_{1}$.
\end{sloppypar}

The flux at $\nu_{m}$ in the Thomson phase for $\nu_{c}<\nu_{m}$ is
\begin{equation}
	F_{\nu_{m}}(t) = 1.2\frac{h_{70}^2}{\left(\sqrt{1+z}-1\right)^2} 
	E_{52}n^{1/2}\epsilon^{1/2}_{B}\left(\frac{\nu_{c}}{\nu_{m}}\right)^{1/2}
	\qquad\mbox{mJy},
	\label{4.12}
\end{equation}
where the numerical coefficient is taken from WG99. For simplicity, the 
cosmological parameters 
have been chosen as $\Omega = 1$ and $\Lambda = 0$ and $h_{70}$ is the 
Hubble constant normalized to $70$ km s$^{-1}$ Mpc$^{-1}$. Since it is 
assumed that second order Compton scattering affects the cooling 
at $\nu_{m}$ only for $t \gtrsim t_{2}$, equation (\ref{4.12}) can be used 
together with the reddening corrected (WG99) value $F_{\nu_{m}}(t_{2}) 
=30$ $\mu$Jy to deduce
\begin{equation}
	E_{52}=3.2 \epsilon_{e}^3,
	\label{4.13}
\end{equation}
or from equation (\ref{4.10})
\begin{equation}
	\epsilon_{B}\epsilon_{e}^7 =8.8\times 10^{-4},
	\label{4.14}
\end{equation}
where $h_{70}=1$ has been used.

During the Thomson phase, the synchrotron energy density peaks 
around $\nu_{m}$ (for $\nu_{c}<\nu_{m}$), while the energy density of 
first order Compton scattered radiation reaches a maximum around 
$\gamma^2_{m}\nu_{m}$. Hence, equation (\ref{2.3}) implies
\begin{equation}
	\nu F_{\nu}(\nu=\nu_{m}) \sim 
	\left(\frac{\epsilon_{B}}{\epsilon_{e}}\right)^{1/2} 
	\nu F_{\nu}(\nu=\gamma^2_{m}\nu_{m}).
	\label{4.15}
\end{equation}
Since cooling at $t_{1}$ due to first and second order Compton 
scattering is assumed to be equal for electrons with Lorentz factors 
around $\gamma_{K}$, at this time the photon energy densities at 
$\nu_{m}$ and $\nu_{KN}(\gamma_{K})$ should be roughly equal. 
Furthermore, in order for the Compton cooling below $\nu_{m}$ at 
$t_{2}$ to occur on the 
$\nu^{1/3}$ part of the first order Compton flux, 
$\nu_{KN}(\gamma_{K})$ cannot be significantly larger than 
$\gamma_{c}^2\nu_{c}$ at $t_{1}$. Hence, an upper limit to 
$\epsilon_{B}/\epsilon_{e}$ is obtained from equation (\ref{4.15}) by 
using $\nu_{KN}(\gamma_{K}) \sim \gamma_{c}^2\nu_{c}$,
\begin{equation}
	\left(\frac{\epsilon_{B}}{\epsilon_{e}}\right)^{1/2}\lesssim 
	\left(\frac{\gamma^2_{c}\nu_{c}}{\gamma^2_{m}\nu_{m}}\right)^{1/2} 
	\sim \frac{\nu_{c}}{\nu_{m}},
	\label{4.16}
\end{equation}
where logarithmic factors in the Compton scattered component 
have been neglected, i.e., a spectral index of $1/2$ is assumed 
between $\gamma^2_{c}\nu_{c}$ and $\gamma_{m}^2\nu_{m}$. From 
equations (\ref{4.11}) and (\ref{4.14}) this results in
\begin{equation}
	\frac{\epsilon^5_{e}}{n}\gtrsim 0.7.
	\label{4.17}
\end{equation}

The major uncertainty in the numerical value in equation (\ref{4.17}) is likely
due to the neglect of light travel time effects. The typical time
for escape through the shockfront for a photon produced in the 
shock is the dynamical time scale. Inclusion of a finite escape time 
for the photons has two major consequences, both of which tend to 
decrease the lower limit in equation (\ref{4.17}). Since cooling is important, 
higher frequencies are produced closer to the shock front than are the 
lower ones. Hence, at a given time, the observed radiation at lower 
frequencies reflects conditions at an earlier time than do that at 
larger frequencies. The rate of energy injection at the shock front 
decreases with time, which suggests the energy density in 
the first Compton component to increase less rapidly between
$\gamma_{c}^2\nu_{c}$ and $\gamma_{m}^2\nu_{m}$ than assumed in 
equation (\ref{4.16}). This expectation is consistent with the 
finding of \citet{Dal00} that the X-ray spectral index as measured by 
BeppoSAX is likely to 
be larger than $0.5$; their preferred value is $0.6$. Furthermore, higher Compton 
orders involve photons which were originally produced at earlier 
times. Again, due to the decreasing rate of energy injection at the 
shockfront, at a given time the relative strength of higher 
Compton orders is increased for larger escape times.
A quantitative evaluation of these effects 
requires numerical calculations. However, it is not likely that they will 
change the main conclusion of the above analysis; namely, in order for
Compton scattering to give a consistent description of the afterglow of GRB 
971214, the value of $\epsilon_{e}$ cannot be significantly below 
unity while that of $n$ cannot be much larger than unity. We note that 
the values deduced for the source parameters of GRB 971214 are rather 
similar to those for GRB 980923; in particular, the value 
of $\epsilon_{e}$ is quite large, which suggests that radiative 
cooling may be important.

The parameter values derived above relied on the assumption that the 
source was in the Thomson phase at $t_{1}$. At $t_{2}$ this is not so 
since the flat or slightly rising spectrum up to $\nu_{m}$ is ascribed
to the effects of second order 
Compton scattering. As emphasized before, inclusion of multiple 
Compton scattering considerably increases the complexity of the 
problem; however, the spectrum at $t_{2}$ presented by WG99 
can be used to make a independent consistency check of 
the scenario discussed above. For the redshift of GRB 
971214, $\nu_{KN}(\gamma_{m}) \approx 10^{17}$ Hz. Since cooling due to 
second order Compton scattering is assumed to be important below 
$\nu_{m}$ at that time, the energy densities at $\nu_{m}$ and 
$\nu_{KN}(\gamma_{m})$ should be approximately equal. Observations show 
that $\nu F_{\nu}$ is roughly a factor of three larger at X-ray 
frequencies ($\approx 10^{18}$) than at $\nu_{m}$. Although the 
spectral shape of the Compton scattered radiation is unknown below the 
X-ray band, this factor may be too small to account for the observed 
synchrotron spectrum below $\nu_{m}$. There are 
two effects, which can account for this possible difference. The reddening 
correction applied by WG99 is substantial. It was 
calculated assuming a pure synchrotron model. However, the 
reddening correction may be smaller, since the 
correlation between intrinsic spectral index and rate of decline of the 
flux, used by WG99, is no longer valid for the scenario 
discussed here. Furthermore, the scattered photons have on average a 
longer escape time than the synchrotron photons. For a given observed flux, 
this results in a larger effective energy density behind the shock of
the former as compared to the latter.

\section{Conclusions}

In the standard model for the afterglows of gamma-ray bursts, the 
observed properties are determined by the values of four parameters 
($E, n, \epsilon_{e}$, and $\epsilon_{B}$). Plausible values of 
$E$ and $n$ come from the overall scenario for the gamma-ray burst 
phenomenon, while the values of $\epsilon_{e}$ and 
$\epsilon_{B}$ are most likely the result of not-so-well understood 
local plasma processes. Although energy requirements favor values of 
$\epsilon_{e}$ not substantially below unity, the value of 
$\epsilon_{B}$ 
is considerably harder to constrain. Equipartition arguments have been 
invoked to motivate the use of $\epsilon_{B}\sim\epsilon_{e}$.

Modelfitting of well observed afterglows can be used to derive values 
of the source parameters. However, for values 
of $E$ and $n$ thought relevant for the standard model, a large 
fraction of the afterglows are likely to have been observed during a phase in 
their evolution in which they suffer from a spectral degeneracy; namely, the 
observed spectra can be fitted equally well by cooling either due to 
synchrotron radiation or Compton scattering. Although, in principle, 
well sampled light curves can break this degeneracy, observations are 
usually not extensive enough to accomplish this. It was argued that the 
afterglows of GRB 980923 and GRB 971214 were observed outside this 
degeneracy phase. The derived source parameters indicate Compton 
cooling to have played an important role during their evolution and, 
hence, $\epsilon_{B}\ll\epsilon_{e}$ at least for these afterglows.

When cooling is enhanced by significant Compton scattering, a 
minor fraction of the radiated energy escapes as synchrotron emission. 
As a result, a given set of observations of the synchrotron component
corresponds, on average, to larger values 
of $\epsilon_{e}$ and/or $E$ than for negligible Compton scattering. 
Although the accuracy of the values determined for the source parameters 
in the afterglows of GRB 980923 and GRB 971214 is rather low, it was 
emphasized that values 
of $\epsilon_{e}$ close to unity are implied for both of them. 
Together with the extended 
times during which the cooling was complete, this suggests the 
possibility of the shock becoming radiative. If cooling in the 
afterglows of most gamma-ray bursts is dominated by Compton 
scattering, the values of $\epsilon_{e}$ would be larger than 
normally deduced. Hence, for a radiative shock, rather small variations in the values of 
$\epsilon_{e}$ could lead to large changes in the emitted flux during 
the later stages of the evolution, for example, when the synchrotron 
emission falls in 
the infrared/optical spectral range. It was also pointed out that in order 
for prompt afterglows, like the one connected to GRB 980923, to be detected by BATSE the 
values of $\epsilon_{e}$ and $\epsilon_{B}$ are quite restricted.

\acknowledgments

This research was supported by a grant from the Swedish Natural 
Science Research Council.

\clearpage

\figcaption[fig1.eps]{The change of dominant cooling mode with time 
for $\epsilon_{B}<\epsilon_{e}$. The different cooling regimes are 
shown for $E_{52}=10$, $n=1$, $\epsilon_{e}=1/2$, and $p=5/2$ 
(synchrotron phase: light shading; Thomson phase: no shading; and multiple 
Compton scattering phase: dark shading). The 
change which occurs for other parameter values is due mainly to 
variations in $t_{eq}$ ($\propto E_{52}n$). The transition between 
different cooling modes is smooth due to Compton scattering in the 
Klein-Nishina limit (see text). \label{fig1}}

\figcaption[fig2.eps]{A schematic representation of the synchrotron 
spectrum in the synchrotron phase when cooling by Compton 
scattering in the Klein-Nishina limit is important for $\eta_{crit}<1$. 
The main features to note are the spectral indices around 
$\nu_{m}$; the one below $\nu_{m}$ is zero, while that in 
the region above $\nu_{m}$ is the same as for negligible cooling. For 
frequencies larger than $\nu_{1}\equiv \eta_{1}^2 \nu_{m}$, cooling 
occurs on photons from the $\nu^{1/3}$-part of the spectrum.
 Effects due to synchrotron cooling and 
synchrotron self-absorption are not included. The spectral breaks 
between the different power-law regimes would be smooth in a real 
spectrum. 
\label{fig2}}

\figcaption[fig3a.eps and fig3b.eps]{Light curves in the Thomson 
phase. Only those breaks are shown which are not present for 
negligible Compton cooling. The break which occurs at $t=t_{eq}$ for 
$\nu>\nu_{eq}$ is achromatic, while the time for the break taking 
place at $t>t_{eq}$ is frequency dependent. Panel (a) corresponds to 
$2<p<8/3$ and panel (b) to $8/3<p<3$. \label{fig3}}

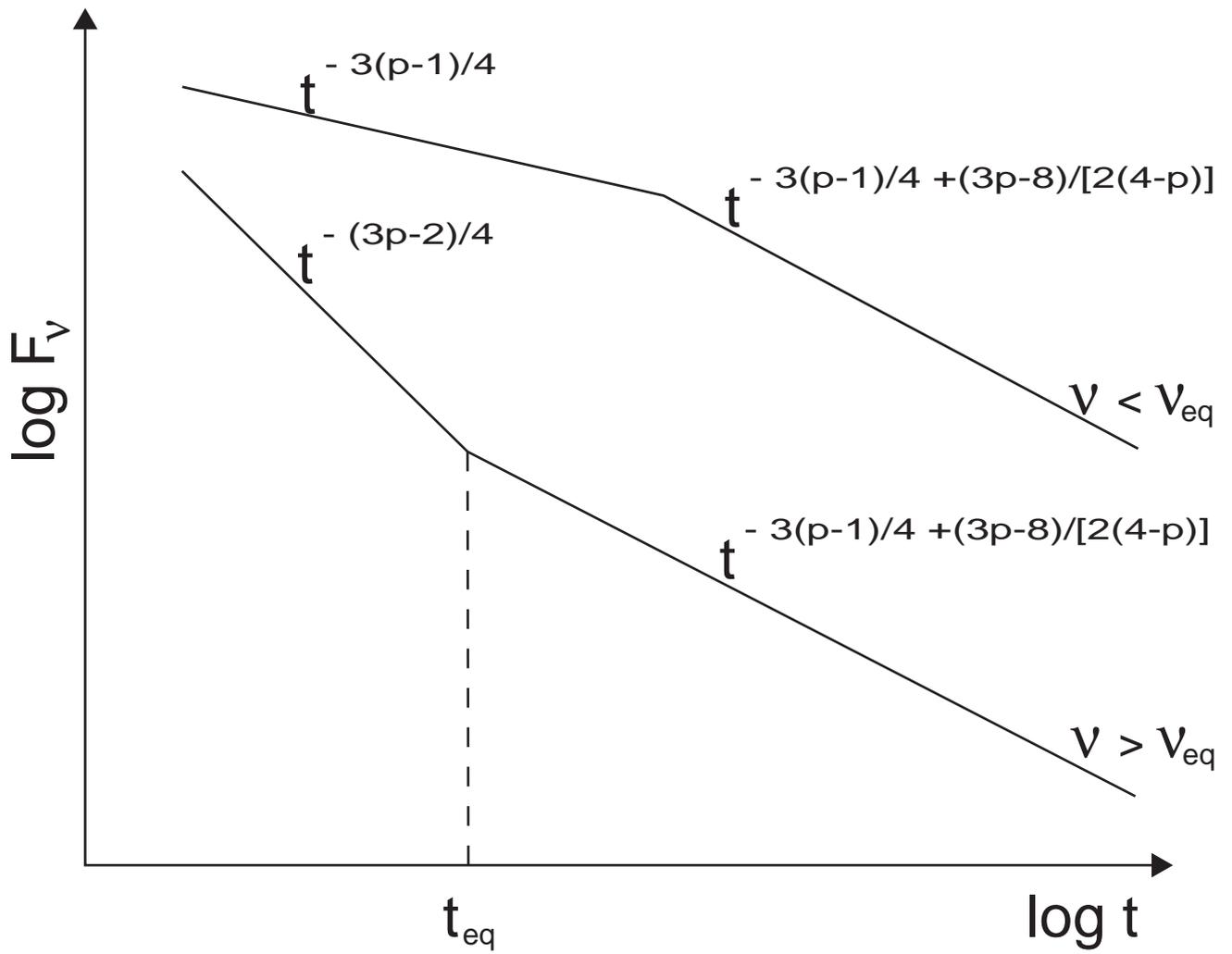
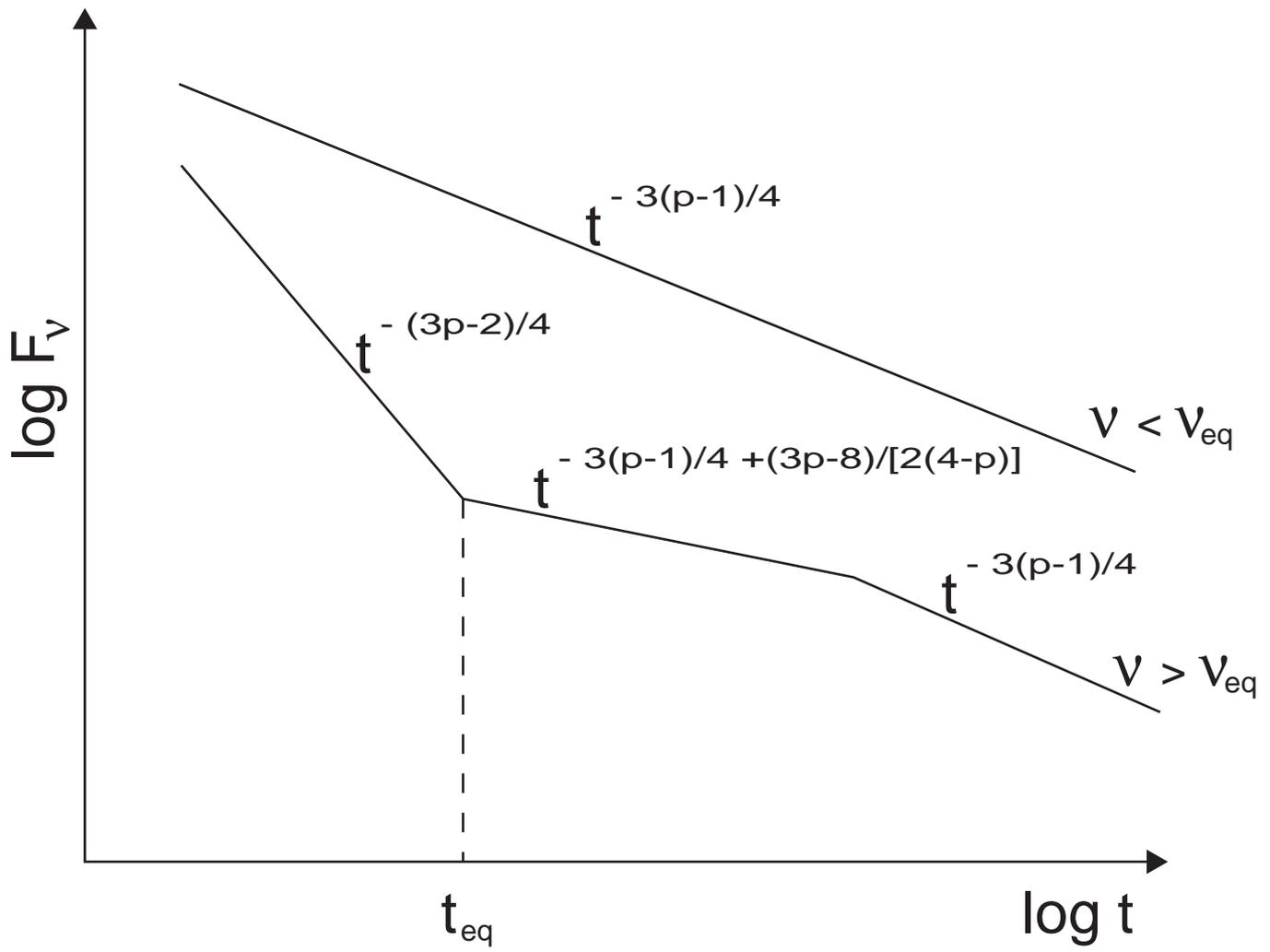
\end{document}